\documentclass[aps,preprint]{revtex4}
\usepackage{graphicx,amsmath,multirow,amssymb}

\draft
\begin{document}
\author{Bin Wang$^{a}$, Zhu-Fang Cui$^{a,c}$, Wei-Min Sun$^{a,b,c}$ and Hong-Shi Zong$^{a,b,c}$}
\address{$^{a}$ Department of Physics, Nanjing University, Nanjing 210093, P. R. China}
\address{$^{b}$ Joint Center for Particle, Nuclear Physics and Cosmology, Nanjing 210093, China}
\address{$^{c}$ State Key Laboratory of Theoretical Physics, Institute of Theoretical Physics, CAS, Beijing 100190, China}
\title{A model study of the chiral phase diagram of QCD }
\begin{abstract}
 In this paper we study the chiral phase transition of QCD at finite temperature and density by using the rank-2 confining separable gluon propagator model in the framework of Dyson-Schwinger Equations. The critical end point is located at $(T_{CEP},\mu_{CEP})=(69~\mathrm{MeV},270.3~\mathrm{MeV})$. It is also found that the first order phase transition might not end at one point, but experiences a two-phase coexisting meta-stable state. A comparison with the results in the previous literature is given.
\bigskip

Key-words: critical end point, chiral susceptibility, chiral phase diagram

\bigskip
 E-mail: zonghs@chenwang.nju.edu.cn.

\bigskip

PACS Numbers: 25.75.Nq, 11.30.Rd, 11.15.Tk

\end{abstract}
\maketitle

\section{Introduction}
It is generally accepted that the chiral symmetry breaking and confining normal hadronic matter will traverse into chiral symmetry restored and deconfining hot dense quark matter (QGP) at high temperature or under high density. It is conjectured that
this new matter exists in the early universe \cite{a1} and the interior of neutron stars \cite{a2,a3,a4,b4,c4}. This new matter can also be created in high energy heavy ion colliders.  One of the main purposes of LHC is to create QGP, probe its properties and chart the phase diagram of QCD. Physicists have spent tens of years to try to plot the phase diagram from theoretical and numerical analysis. A prevalent viewpoint is that this diagram has a critical end point which connects a crossover line for higher temperature and lower chemical potential and a first order transition line for lower temperature and higher chemical potential \cite{a5,a6,a7,a8}. Charting the phase diagram, testing the existence of CEP and locating its position is one of the most active field in high energy physics. Benefiting from the improvement of computer technology, lattice QCD has been the most important means of studying non-perturbative QCD. Many works based on this tool indicate the existence and give the position of CEP \cite{a10,l1,l2,l3}. But lattice QCD still cannot give a convincing result because of the notorious fermion sign problem, so the calculations based on effective theories of QCD are also irreplaceable nowadays. For example, the NJL model is used in many works on chiral phase transition \cite{a111,a11,b111,b1,c1}. The existence of CEP is also shown and its probable position is estimated, too. But among these works, someone gives the phase diagram not completely consistent with the popular viewpoint, i. e. a two phase coexisting domain might replace the first order transition line \cite{b1}.

The Dyson-Schwinger Equations (DSE) is another continuum approach, which has been proved to be a useful theoretical tool in the study related to confinement and dynamical chiral symmetry breaking.
The study of the chiral phase diagram with this tool is also present in recent years \cite{a12,b12,c12,a13,b2}. All of them indicate the existence of the CEP and some papers also report the appearance of the two phase coexisting domain \cite{a13,b2}. However, in Ref. \cite{a13} only the case of chiral limit is investigated, while in Ref. \cite{b2} the authors have investigated the case of finite current quark mass within a truncation scheme including the temperature and in-medium effects of the gluon propagator. In this paper we try to plot the chiral phase diagram  using the DSE approach by extrapolating a generally used model, i.e. the rank-2 confining separable model gluon propagator, to finite temperature and finite chemical potential. Our study gives the position of CEP and shows explicitly that a two phase coexisting domain also appears beyond the chiral limit.

\section{Theoretical and Numerical Analysis}
\subsection{the chiral phase transition at finite temperature and $\mu=0$}

To be self-contained, let us first give a brief introduction to
the DSE. The Dyson-Schwinger Equations consist of an infinite tower of
integral equations correlated with each other. In fact it cannot be
solved exactly. Anyone intending to give a solution of it must do
some truncation to break the chains of the infinite tower. There are
many schemes on this in the literature, such as the quench
approximation, the rainbow-ladder approximation, etc. (for review
articles, see Refs. \cite{a14,a15}). In the rainbow approximation, the gap equation at
finite temperature can be written as
\begin{equation}
G^{-1}(p_k)=i\gamma\cdot p_k+m+\frac{4}{3}T\sum_{n=-\infty}^{+\infty}
\int\frac{d^3q}{(2\pi)^3}g^2D^{eff}_{\mu\nu}(p_k-q_n)\gamma_{\mu}G(q_n)\gamma_{\nu}.
\end{equation}
in which the inverse of the quark propagator
$G^{-1}(p_k)$ can be decomposed as
\begin{equation}
G^{-1}(p_k)=i\vec{\gamma}\cdot\vec{p}A(p_k^2)+i\gamma_4\omega_k C(p_k^2)+B(p_k^2)
\end{equation}
and $D_{\mu\nu}^{eff}(p_k-q_n)$ denotes the effective gluon propagator. In nowadays literature of DSE, the effective model gluon propagator is often introduced as a physical input, and the quark propagator is calculated out by the gap equation with this input. There are two qualitative requirements for the effective gluon propagator in DSE approach. First, the effective gluon propagator should simulates the infrared enhancement and confinement. Second, this gluon propagator should lead to the dynamical chiral symmetry breaking and the obtained quark propagator has no particle-like poles on the timelike $p^2$ axis(so that quarks are confined). In other words, the physical input of the effective gluon propagator must ensure that the DSE has the features of confinement and dynamical chiral symmetry breaking simultaneously.
The rank-2 confining separable model gluon propagator is a generally used effective model in the literature which was first proposed for describing the properties of light flavor pseudo-scalar and vector mesons \cite{a16,a17}.
At finite temperature, this model gluon propagator can be written as:
\begin{equation}
g^2D^{eff}_{\mu\nu}(p_k-q_n)=\delta_{\mu\nu}[D_0f_0(p^2_k)f_0(q^2_n)+D_1f_1(p_k^2)p_k\cdot q_n f_1({q}_n^2)],
\end{equation}
in which ${q}_n=(\vec{q},{\omega}_n)$, with ${\omega}_n=(2n+1)\pi T$ \cite{a17,a18,b18},
$f_0({q}_n^2)=\exp(-{q}_n^2/\Lambda^2)$, $f_1({q}_n^2)=\exp(-{q}_n^2/\Lambda_1^2)$,$\Lambda_0=0.638~\mathrm{GeV}$, $\Lambda_1/\Lambda_0=1.21$, $D_0\Lambda_0^2=260.0$, $D_1\Lambda_1^4=130.0$ and the degenerated light quark mass $m=5.3 ~\mathrm{MeV}$ \cite{a17}.

Substituting Eq. (2) and Eq. (3) into the gap equation (1), one obtains the following coupled integral equations:
\begin{eqnarray}
A({p}_k^2)&=&1+a(T)f_1({p}_k^2),\\
B({p}_k^2)&=&m+b_0(T)f_0({p}_k^2)+b_1(T){\omega}_kf_1({p}_k^2),\\
C({p}_k^2)&=&1+c_0(T)f_0({p}_k^2)/{\omega}_k+c_1(T)f_1({p}_k^2),
\end{eqnarray}
in which
\begin{eqnarray}
a(T)&=&\frac{8}{3}D_1T\sum_{n=-\infty}^{+\infty}\int\frac{d^3q}{(2\pi)^3}f_0({q}_n^2)\vec{q}^2A({q}_n^2)d^{-1}({q}_n^2),\\
b_0(T)&=&\frac{16}{3}D_0T\sum_{n=-\infty}^{+\infty}\int\frac{d^3q}{(2\pi)^3}f_0({q}_n^2)B({q}_n^2)d^{-1}({q}_n^2),\\
b_1(T)&=&\frac{16}{3}D_1T\sum_{n=-\infty}^{+\infty}\int\frac{d^3q}{(2\pi)^3}f_1({q}_n^2){\omega}_nB({q}_n^2)d^{-1}({q}_n^2),\\
c_0(T)&=&\frac{8}{3}D_0T\sum_{n=-\infty}^{+\infty}\int\frac{d^3q}{(2\pi)^3}f_0({q}_n^2){\omega}_nC({q}_n^2)d^{-1}({q}_n^2),\\
c_1(T)&=&\frac{8}{3}D_1T\sum_{n=-\infty}^{+\infty}\int\frac{d^3q}{(2\pi)^3}f_1({q}_n^2){\omega}^2_nC({q}_n^2)d^{-1}({q}_n^2),
\end{eqnarray}
where $d({p}_k^2)=\vec{p}^2A^2({p}_k^2)+{\omega}_k^2C^2({p}_k^2)+B^2({p}_k^2)$.
\begin{figure}
  \includegraphics[width=12cm]{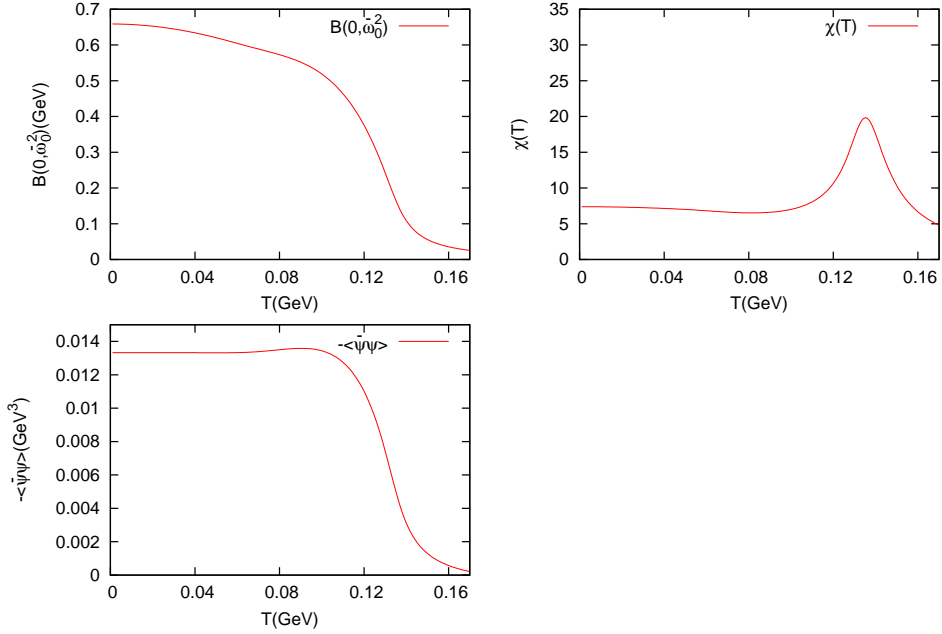}\\
  \caption{The upper left panel is the curve of $B(\vec{0},{\omega_0^2})$ versus $T$. The lower left panel is the curve of $-\langle\overline{\psi}\psi\rangle$ versus $T$. The upper right panel is the curve of $\chi(T)$ versus $T$.}\label{dbdbmtem}
\end{figure}

In the upper left panel of Fig. \ref{dbdbmtem}  we plot the dependence of $B(0,\omega_0^2)$ on $T$. At lower temperature the value of $B(0,\omega_0^2)$ is much bigger than the bare current quark mass $m$, which indicates that chiral symmetry is spontaneously broken. As the temperature increases, the value of $B(0,\omega_0^2)$ becomes nearer and nearer to the bare quark mass $m$, which shows the process of restoration of chiral symmetry.

Before extrapolating this model to finite temperature and chemical potential, its validity in the study of chiral phase transition at finite temperature and $\mu=0$ should be shown at first. In fact, these works have been done in \cite{a17,a18}. But for the readers' convenience, here we would like to calculate some observables, such as the quark condensate and the chiral susceptibility.
The quark condensate at finite temperature and zero chemical potential can be written as
\begin{eqnarray}
\label{qq}
-\langle\overline{\psi}\psi\rangle(T)
=N_c N_fT\sum_{n=-\infty}^{+\infty}\int\frac{d^3q}{(2\pi)^3}tr_\gamma[G(p_n,m)],
\end{eqnarray}
in which $G$ is the quark propagator, $N_c=3$ is the color factor,
$N_f=2$ denotes two degenerate light flavors and the trace operation
is over Dirac indices of the quark propagator.
Here we would note that the quark condensate as defined in Eq.~(\ref{qq}) is divergent, which can be easily seen from its form in the ultraviolet limit
\begin{eqnarray}
\label{qquv1}
-<\overline{\psi}\psi>_{UV}(T)&=&N_c N_fT\sum_{n=-\infty}^{+\infty}\int\frac{d^3q}{(2\pi)^3}tr_\gamma[\frac{1}{i\vec{\gamma}\cdot\vec{q}+i\gamma_4{\omega_n}+m}]\nonumber\\
&=&-\frac{N_c N_f}{\pi^2}\int_0^{\infty}dq\,\big(\frac{q^2m}{w}[\frac{1}{1+e^{(\omega-\mu)/T}}+\frac{1}{1+e^{(\omega+\mu)/T}}]-\frac{q^2m}{w}\big)
\label{qquv2}
\end{eqnarray}
in which $w=(q^2+m^2)^{1/2}$. The last term in Eq. (\ref{qquv2}) is divergent but it does not depends on $T$. Since we only concern about the variation of the chiral
susceptibility with respect to $T$, this term can be
dropped safely. In order to eliminate the ultraviolet divergence in Eq. (13), we will subtract this divergent term from the right-hand side of Eq. (12).

The chiral susceptibility is defined as the derivative of the chiral order parameter with respect to current quark mass, one generally used definition in the literature of DSE is \cite{a13,a15,d,a19,a21}:
\begin{eqnarray}
\label{chi}
\chi(T)=\frac{\partial}{\partial m}B(\vec{0},{\omega_0^2}).
\end{eqnarray}
This definition only includes the derivative of the B-function evaluated at zero-momentum and for the lowest Matsubara frequency and does not consider the contribution of all the rest of the modes. Because $B(\vec{0},{\omega_0^2})$ can completely determine the character of the chiral phase transition, so that it is a bona fide order parameter as the quark condensate is and the definition of Eq. (\ref{chi}) is equivalent to that defined as the derivative of quark condensate with respect to current quark mass \cite{d,a19,b19}.  We will also show this equivalence numerically below.

To  obtain $\frac{\partial}{\partial m}B(\vec{0},{\omega_0^2})$, the derivative of $A({p}_k^2)$, $B({p}_k^2)$ and $C({p}_k^2)$ with respect to the current quark mass is needed:
\begin{eqnarray}
A_m(p_k^2)&=&a_m(T)f_1({p}_k^2),\\
B_m(p_k^2)&=&1+b_{0m}(T)f_0({p}_k^2)+b_{1m}(T){\omega}_kf_1({p}_k^2),\\
C_m(p_k^2)&=&c_{0m}(T)f_0({p}_k^2)/{\omega}_k+c_{1m}(T)f_1({p}_k^2),
\end{eqnarray}

in which $A_m(p_k^2)=\frac{\partial A({p}_k^2)}{\partial m}$,$B_m(p_k^2)=\frac{\partial B({p}_k^2)}{\partial m}$,$C_m(p_k^2)=\frac{\partial C({p}_k^2)}{\partial m}$, and
\begin{eqnarray}
a_m(T)&=&\frac{8}{3}D_1T\sum_{n=-\infty}^{+\infty}\int\frac{d^3q}{(2\pi)^3}f_0({q}_n^2)\vec{q}^2[A_m({q}_n^2)d^{-1}({q}_n^2)-A({q}_n^2)d_m({q}_n^2)d^{-2}({q}_n^2)],\\
b_{0m}(T)&=&\frac{16}{3}D_0T\sum_{n=-\infty}^{+\infty}\int\frac{d^3q}{(2\pi)^3}f_0({q}_n^2)[B_m({q}_n^2)d^{-1}({q}_n^2)-B({q}_n^2)d_m({q}_n^2)d^{-2}({q}_n^2)],\\
b_{1m}(T)&=&\frac{16}{3}D_1T\sum_{n=-\infty}^{+\infty}\int\frac{d^3q}{(2\pi)^3}f_1({q}_n^2){\omega}_n[B_m({q}_n^2)d^{-1}({q}_n^2)-B({q}_n^2)d_m({q}_n^2)d^{-2}({q}_n^2)],\\
c_{0m}(T)&=&\frac{8}{3}D_0T\sum_{n=-\infty}^{+\infty}\int\frac{d^3q}{(2\pi)^3}f_0({q}_n^2){\omega}_n[C_m({q}_n^2)d^{-1}({q}_n^2)-C({q}_n^2)d_m({q}_n^2)d^{-2}({q}_n^2)],\\
c_{1m}(T)&=&\frac{8}{3}D_1T\sum_{n=-\infty}^{+\infty}\int\frac{d^3q}{(2\pi)^3}f_1({q}_n^2){\omega}^2_n[C_m({q}_n^2)d^{-1}({q}_n^2)-C({q}_n^2)d_m({q}_n^2)d^{-2}({q}_n^2)]),
\end{eqnarray}
where $d_m({p}_k^2)={\partial d({p}_k^2)}/{\partial m}=
2\vec{p}^2A({p}_k^2)\,A_m({p}_k^2)+2{\omega}_k^2C({p}_k^2)\,C_m({p}_k^2)+2B({p}_k^2)\,B_m({p}_k^2)$, in which $a_m(T)=\frac{\partial a(T)}{\partial m}$,$b_{0m}(T)=\frac{\partial b_0(T)}{\partial m}$,$b_{1m}(T)=\frac{\partial b_1(T)}{\partial m}$,$c_{0m}(T)=\frac{\partial c_0(T)}{\partial m}$,$c_{1m}(T)=\frac{\partial c_1(T)}{\partial m}$.

The dependence of the quark condensate and the chiral susceptibility on $T$ is shown in the lower left and upper right panels of Fig. \ref{dbdbmtem}, respectively. It can be found that the quark condensate undergoes a continuously dramatic change in the neighborhood of $T=136~\mathrm{MeV}$. The chiral susceptibility changes continuously in the whole temperature range and a peak presents at $T=136~\mathrm{MeV}$. These behaviors imply that the chiral phase transition at finite temperature and zero chemical potential is a crossover.
Although there are some defects caused by the artefact of the separable model, e.g., the numerical results show that in the range $52~\mathrm{MeV}<T<90~\mathrm{MeV}$ the quark condensate increases as the temperature increases (in fact, it should be monotonously decreasing) and the location of the peak of chiral susceptibility is relatively smaller compared to lattice result \cite{lattice1}, this model does give results qualitatively consistent with nowadays prevailing viewpoint \cite{a17,a18}.
Simplicity is a big advantage of this model, it overcomes the difficulty in the summation of the frequency spectrum at finite temperature confronted in many other more sophisticated models \cite{d1,d2,d3} but it can be used to highlight many important underlying mechanisms.
\subsection{the chiral phase transition at finite temperature and finite chemical potential}
As the rank-2 separable model shows its validity in the study of chiral phase transition at finite temperature, we hope this model can be extrapolated to the case with both nonzero density and nonzero temperature. Since the chemical potential $\mu$ enters into the gluon propagator via the Debye screening effect, if we don't consider this effect as it is at zero chemical potential and finite temperature (see Eq. (3)), it's form will still be the same
\begin{equation}
\label{sepTmu}
g^2D^{eff}_{\mu\nu}(\widetilde{p}_k-\widetilde{q}_n)=\delta_{\mu\nu}[D_0f_0(p^2_k)f_0(q^2_n)+D_1f_1(p_k^2)p_k\cdot q_n f_1({q}_n^2)],
\end{equation}
in which $\widetilde{q}_n=(\vec{q},\widetilde{\omega}_n)$, with $\widetilde{\omega}_n=\omega_n+i\mu$.
When the temperature and the chemical potential are both nonzero, the gap equation, the inverse of the quark propagator and the chiral susceptibility will become
\begin{equation}
G^{-1}(\widetilde{p}_k)=i\gamma\cdot\widetilde{p}_k+m+\frac{4}{3}T\sum_{n=-\infty}^{+\infty}
\int\frac{d^3q}{(2\pi)^3}g^2D^{eff}_{\mu\nu}(\widetilde{p}_k-\widetilde{q}_n)\gamma_{\mu}G(\widetilde{q}_n)\gamma_{\nu},
\end{equation}
\begin{equation}
G^{-1}(\widetilde{p}_k)=i\vec{\gamma}\cdot\vec{p}A(\widetilde{p}_k^2)+i\gamma_4\widetilde{\omega_k}C(\widetilde{p}_k^2)+B(\widetilde{p}_k^2).
\end{equation}
\begin{eqnarray}
\chi(T,\mu)=\frac{\partial}{\partial m}B(\vec{0},\widetilde{\omega_0}^2).
\end{eqnarray}
Following the same procedure in the finite temperature case, we would obtain the analogous equations corresponding to Eq.~(4)-(11) and Eq.~(15)-(22).

\begin{figure}
  \includegraphics[width=12cm]{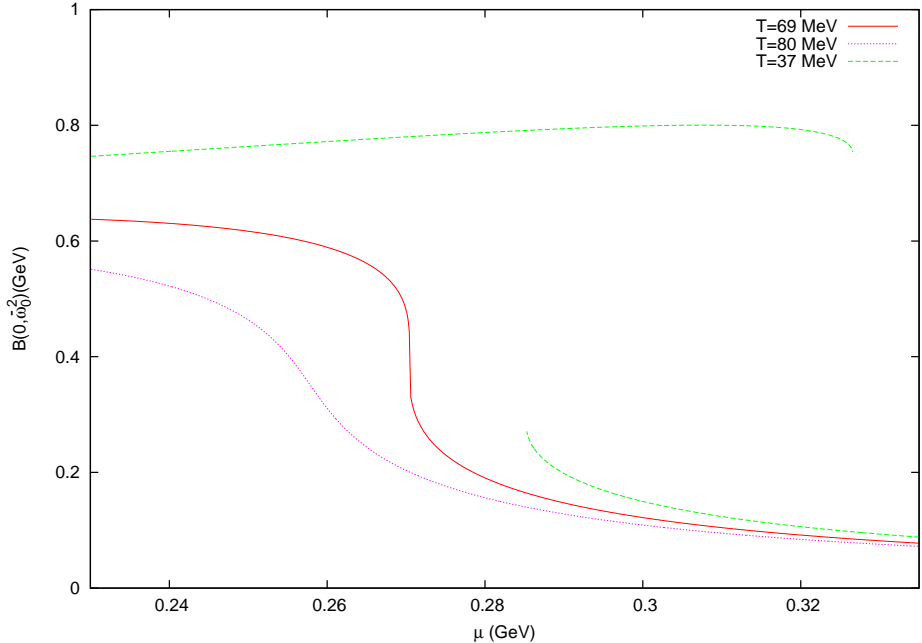}\\
  \caption{Dependence of $B(0,\widetilde{\omega}_0^2)$ on $\mu$ for $T$ equal to $80~\mathrm{MeV}$, $69~\mathrm{MeV}$ and $37~\mathrm{MeV}$, respectively.}\label{Fig1}
\end{figure}

In Fig. \ref{Fig1} we plot the dependence of $B(0,\widetilde{\omega}_0^2)$ on $\mu$ for $T$ equal to $80~\mathrm{MeV}$, $69~\mathrm{MeV}$ and $37~\mathrm{MeV}$, respectively. When the temperature is equal to $80~\mathrm{MeV}$ and $69~\mathrm{MeV}$, $B(0,\widetilde{\omega}_0^2)$ has only one solution and changes continuously from the Nambu solution (the larger value) to the Wigner solution (the smaller value). But in the transition domain the line for $T=69~\mathrm{MeV}$ is much steeper than that for $T=80~\mathrm{MeV}$. For the case $T=37~\mathrm{MeV}$, when $\mu$ is between $\mu_{W}=285.2~\mathrm{MeV}$ and $\mu_N=326.6~\mathrm{MeV}$, $B(0,\widetilde{\omega}_0^2)$ has both Nambu solution and Wigner solution; it has only one Nambu solution when $\mu<\mu_W$ and has only one Wigner solution when $\mu>\mu_N$. Each jump point appears at the endpoint of the Nambu solution $\mu_N$ and the endpoint of the Wigner solution $\mu_W$ respectively, this behavior has been previously recognized as a signal for phase coexistence in the literature \cite{a13,b1,b2} and we will discuss this problem further below.

\begin{figure}
  \includegraphics[width=12cm]{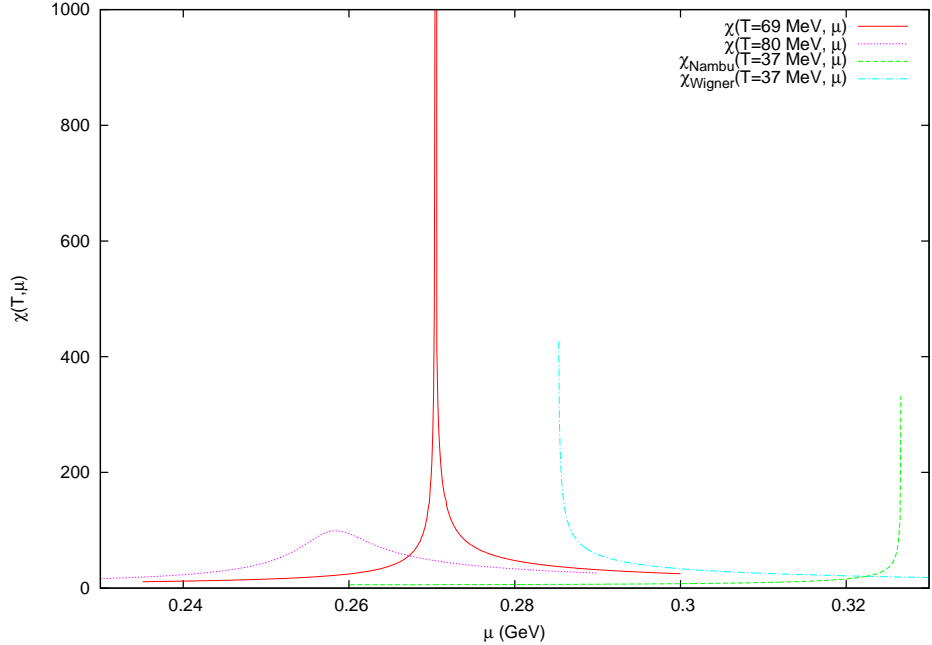}\\
  \caption{Dependence of the chiral susceptibility $\chi(T,\mu)$ on $\mu$ for $T$ equal to $80~\mathrm{MeV}$, $69~\mathrm{MeV}$ and $37~\mathrm{MeV}$, respectively.}\label{Fig2}
\end{figure}
In Fig. \ref{Fig2} we plot the dependence of the chiral susceptibility $\chi(T,\mu)$ on $\mu$ for $T$ equal to $80~\mathrm{MeV}$, $69~\mathrm{MeV}$ and $37~\mathrm{MeV}$, respectively. The most prominent one is that for $T=69~\mathrm{MeV}$, it is divergent at $\mu=270.3~\mathrm{MeV}$ which means a second order phase transition occurs at this point (in fact this point is the CEP). At the temperature higher than $69~\mathrm{MeV}$ the chiral susceptibility $\chi(T,\mu)$ would change continuously and a peak with finite height appears at one point which means that the chiral phase transition would happen as a crossover there. But at the temperature smaller than $69~\mathrm{MeV}$ the chiral susceptibility $\chi(T,\mu)$ changes discontinuously while each peak and jump point appear at $\mu_W$ and $\mu_N$ respectively. It looks like that the chiral phase transition is of first order and has two different transition points at this temperature.

\begin{figure}
  \includegraphics[width=12cm]{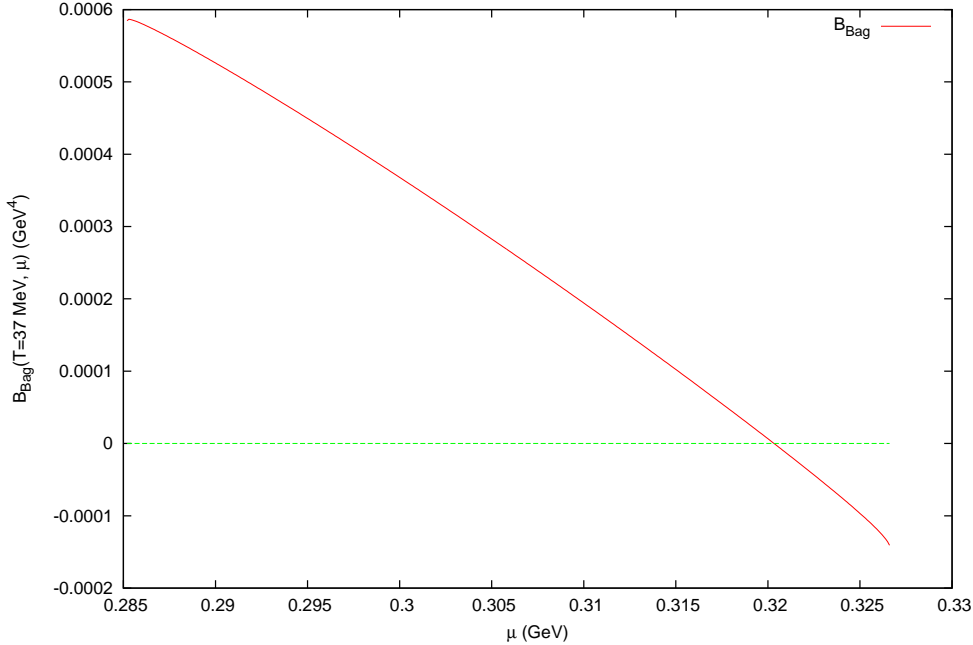}\\
  \caption{The dependence of the bag constant on $\mu$ when $T=37~\mathrm{MeV}$.}\label{Fig3}
\end{figure}
In the literature the bag constant $B_{bag}(T,\mu)$, i.e. the difference between the pressure of the Nambu solution and that of the Wigner solution, is often used to determine which solution is the physical one when both the Nambu solution and the Wigner solution coexist at the same point. The solution with the higher pressure is regarded as the physical one. The Cornwall-Jackiw-Tomboulis (CJT) approximation is the most generally used tool to calculate the pressure, whose finite temperature version can be written as \cite{a20}:
\begin{equation}
P(T,\mu)=T\sum_{n=-\infty}^{+\infty}\int\frac{d^3q}{(2\pi)^3}Tr\biggl\{Ln[G^{-1}(\widetilde{p}_n)G_0(\widetilde{p}_n)]+\frac{1}{2}[G_0^{-1}(\widetilde{p}_n)G(\widetilde{p}_n)-1]\biggr\},
\end{equation}
where $G(\widetilde{p}_n)$ and $G_0(\widetilde{p}_n)$ stand for the full and free quark propagator, respectively. The variation of the bag constant $B_{bag}(T,\mu)$ along with the chemical potential in the range that two solutions coexist is shown in Fig. \ref{Fig3}. For the case $T=37~\mathrm{MeV}$, the bag constant is positive when $\mu$ is smaller than $320.4~\mathrm{MeV}$ and is negative when $\mu$ is larger than it. From this viewpoint the first order chiral phase transition occurs at $\mu_c=320.4~\mathrm{MeV}$. Then, how does one reconcile this conclusion with the one we obtained from the chiral susceptibility in the previous figure? Here, we will take the viewpoint presented in Ref. \cite{a13}. The matter should be in Nambu phase when $\mu<\mu_c$ and in Wigner phase when $\mu>\mu_c$.
But like the superheating phenomenon in thermodynamics, there might exist some mechanism that prevents the chiral phase transition from occurring immediately at $\mu_c$ and there is a meta-stable phase in addition to the stable phase in the neighborhood of $\mu_c$. When $\mu_W<\mu<\mu_c$, the Nambu phase is stable but the Wigner phase is meta-stable. When $\mu_c<\mu<\mu_N$, the Wigner phase is stable but the Nambu phase is meta-stable. If matter is in the meta-stable Nambu phase, pockets of deconfined, chirally symmetric quark matter might appear in the confining Nambu medium because of the fluctuation effect. And there is the analogous phenomenon in meta-stable Wigner phase. So the two phases might coexist when $\mu_W<\mu<\mu_N$, but there is only Nambu phase when $\mu<\mu_W$ and only Wigner phase when $\mu>\mu_N$.

\begin{figure}
  \includegraphics[width=12cm]{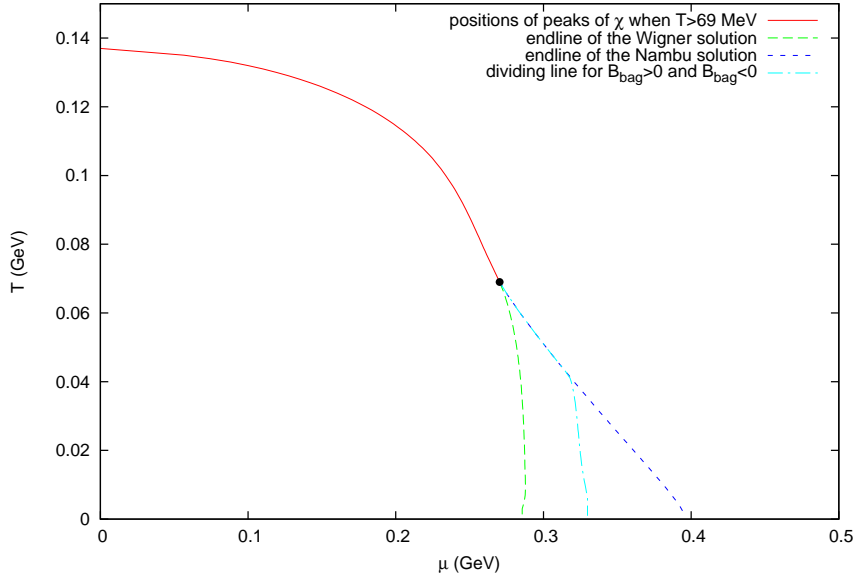}\\
  \caption{The phase diagram (the black point is the CEP)}\label{Fig4}
\end{figure}

The phase diagram is presented in Fig. \ref{Fig4}. The chiral phase transition is a crossover when $T>69~\mathrm{MeV}$ and it is a first order phase transition when $T<69~\mathrm{MeV}$. A CEP exists at $(T_{CEP},\mu_{CEP})=(69~\mathrm{MeV},270.3~\mathrm{MeV})$,  and the ratio $T_{CEP}/\mu_{CEP}$ is equal to $0.255$.
Our ratio is much smaller than the ratio given by more realistic models which is approximately equal to $1$ \cite{a13,a10,l1,l2,l3}, and is even smaller than the result given by NJL-like models \cite{a111,a11,b111,b1}.
The chiral symmetry restored line (it consists of the end points of the Nambu solution and its left side is the chiral symmetry restored phase) and the chiral symmetry broken line  (it consists of the end points of the Wigner solution and its right side is the chiral symmetry broken phase)  bifurcate from the CEP and a two-phase-coexisting domain appears between these two lines (here a two-phase coexisting domain means a certain domain of chemical potential and temperature on which both the Wigner- and Nambu-phase susceptibilities are positive). The border line between the domains with different bag constant sign is also presented. The two phases coexisting domain
between this border line and the chiral symmetry broken line consists of the stable Nambu phase and meta-stable Wigner phase while that between this border line and the chiral symmetry restored line consists of the stable Wigner phase and meta-stable Nambu phase.

\begin{figure}
  \includegraphics[width=12cm]{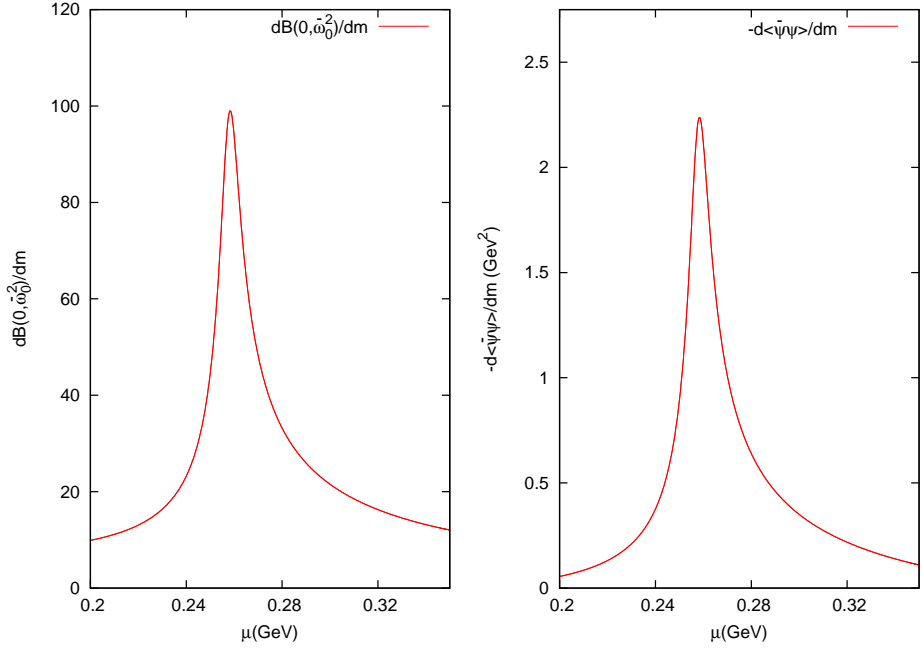}\\
  \caption{The dependence of $\chi(T=80~\mathrm{MeV},\mu)$ (the left panel) and $\chi^{\star}(T=80~\mathrm{MeV},\mu)$ (the right panel) on chemical potential.}\label{Figqqmdbm80}
\end{figure}

Now let us turn back to show the relation numerically between the two definitions of the chiral susceptibility.
Just as was pointed out previously, the chiral susceptibility defined in Eq. (14) only includes the lowest Matsubara frequency but it could give qualitatively consistent result to those obtained from the more conventional definition which include all Matsubara frequencies
\begin{eqnarray}
\chi^{\star}&=&(-)\frac{\partial}{\partial
m}<\overline{\psi}\psi>
=(-)N_cN_fT\sum_{n=-\infty}^{+\infty}\int \frac{d^3 q}{(2 \pi)^3}tr_\gamma[G(\widetilde{q}_n,m)\frac{\partial
G^{-1}(\widetilde{q}_n,m)}{\partial m}G(\widetilde{q}_n,m)].
\end{eqnarray}
Like the quark condensate, this quantity is also ultraviolet divergent which can be seen from its ultraviolet limit form
\begin{eqnarray}
\chi^{\star}_{UV}&=&(-)N_cN_fT\sum_{n=-\infty}^{+\infty}\int \frac{d^3 q}{(2 \pi)^3}
tr_\gamma[\frac{1}{i\vec{\gamma}\cdot\vec{q}+i\gamma_4\widetilde{\omega_n}+m}\frac{1}{i\vec{\gamma}\cdot\vec{q}+i\gamma_4\widetilde{\omega_n}+m}] \\\
&=&-\frac{N_fN_c}{\pi^2}\int_0^\infty dqq^2\biggl\{(\frac{1}{\omega}-\frac{m^2}{\omega^3})[\frac{1}{1+e^{(\omega-\mu)/T}}+\frac{1}{1+e^{(\omega+\mu)/T}}] \nonumber\\
&&-\frac{m^2}{\omega^2T}[\frac{e^{(\omega-\mu)/T}}{[1+e^{(\omega-\mu)/T}]^2}+\frac{e^{(\omega+\mu)/T}}{[1+e^{(\omega+\mu)/T}]^2}]-(\frac{1}{\omega}-\frac{m^2}{\omega^3})\biggr\},
\end{eqnarray}
where $\omega=(q^2+m^2)^{1/2}$. The last term in Eq. (30) will cause the integral divergent and is not related to $\mu$ and $T$. As we only concern about the variation of the chiral
susceptibility with respect to $\mu$ and $T$, this term can be
removed directly. In the numerical calculation of $\chi^{\star}$ we can include all the frequencies with the following trick. Because of asymptotic freedom, the dressed quark propagator will become the free quark propagator at high frequencies, i.e. $G^{-1}(\widetilde{p}_k,m)=i\vec{\gamma}\cdot\vec{p}A(\widetilde{p}_k^2)+i\gamma_4\widetilde{\omega_k}C(\widetilde{p}_k^2)
+B(\widetilde{p}_k^2)\rightarrow i\vec{\gamma}\cdot\vec{p}+i\gamma_4\widetilde{\omega_k}+m$, $\partial G^{-1}(\widetilde{p}_k,m)/\partial m\,\rightarrow 1$.
So that the integrand of Eq. (28) will be the same to that of Eq. (29) at high frequencies. Since the summation of all frequencies in Eq. (29) can be calculated out analytically, so that we can first subtract Eq. (29) from Eq. (28) and then plus Eq. (30). In our actual numerical calculation, the summation will be terminated at high enough frequency beyond which the integrand of Eq. (28) and Eq. (29) is equal (specifically our calculation terminates at $\widetilde{\omega_n}^2\ge 15*\Lambda_1$, which ensures that our numerical result does not depend on the choice of the cutoff in the Matsubara frequencies).

For comparison, we plot the curves of $\chi(T=80~\mathrm{MeV},\mu)$ and $\chi^{\star}(T=80~\mathrm{MeV},\mu)$ in Fig. \ref{Figqqmdbm80}. The shapes of these two curves are analogous, and the positions of the peaks of these two curves are very near to each other ($\mu=258.4~\mathrm{MeV}$ for $\chi(T=80~\mathrm{MeV},\mu)$, $\mu=258.5~\mathrm{MeV}$ for $\chi^{\star}(T=80~\mathrm{MeV},\mu)$). This consistence is what we have expected and is the reason why we choose Eq. (14) in this work
as the definition of chiral susceptibility and draw our conclusion from the analysis of this quantity.
\section{SUMMARY AND CONCLUSIONS}
The destination of this paper is to plot the QCD phase diagram and locate the critical end point (CEP) in the framework of the Dyson-Schwinger Equation. To this end we first pin down the form of the rank-2 confining separable model gluon propagator at nozero chemical potential and finite temperature which is the same to that at zero chemical potential and finite temperature. By solving the gap equation we find that the Nambu solution changes continuously to the Wigner solution as the chemical potential increases when $T>69~\mathrm{MeV}$ but changes discontinuously and has a two solution coexisting domain when $T<69~\mathrm{MeV}$. Then we use the chiral susceptibility and the bag constant to study the chiral phase transition. Our research indicates that a critical end point exists at $(T_{CEP},\mu_{CEP})=(69~\mathrm{MeV},270.3~\mathrm{MeV})$, with the ratio $T_{CEP}/\mu_{CEP}=0.255$ which is a strongly model dependent result and is much smaller than the ratio $T_{CEP}/\mu_{CEP}\thickapprox1$  given  by the lattice QCD and the DSE with more sophisticated model  \cite{a13,a10,l1,l2,l3}. Our work also shows that the first order phase transition might not end at one point but experiences a two phases coexisting meta-stable state.

\acknowledgments

This work is supported in part by the National Natural Science Foundation of China (under Grant 11275097, 10935001 and 11075075) and the Research Fund for the Doctoral Program of Higher Education (under Grant No 2012009111002).

\end{document}